\documentclass[letterpaper]{article}
\usepackage{graphicx}

\begin{document}
\date {}
\title{ THE MAJORANA $^{76}$Ge DOUBLE-BETA DECAY
PROJECT  }

\footnotetext[1]{Pacific Northwest National Laboratory, Richland,
WA 99352, USA}
\footnotetext[2]{University of South Carolina, Columbia, SC
29208, USA}
\footnotetext[3]{Institute for Theoretical and Experimental
Physics, Moscow 117259, Russia}
\footnotetext[4]{Joint Institute for Nuclear Research, Dubna,
Russia}
\footnotetext[5]{University of Chicago, Chicago, IL, USA}
\footnotetext[6]{University of Washington, Seattle, WA, USA}
\footnotetext[7]{Brown University, Providence, RI, USA}
\footnotetext[8]{Triangle Universities Nuclear Laboratory, Durham,
NC, USA}
\footnotetext[9]{Duke University, Durham, NC, USA}
\footnotetext[10]{New Mexico State University, NM, USA}
\footnotetext[11]{North Carolina State University, Raleigh, NC ,
USA}

\maketitle

\author{C.E. Aalseth\footnotemark[1],
D. Anderson\footnotemark[1],
F.T. Avignone III\footnotemark[2],
A. Barabash\footnotemark[3],
T.W. Bowyer\footnotemark[1],
R.L. Brodzinski\footnotemark[1],
V. Brudanin\footnotemark[4],
J.I. Collar\footnotemark[5],
P.J. Doe\footnotemark[6],
S. Egorov\footnotemark[4],
S.R. Elliott\footnotemark[6],
H.A. Farach\footnotemark[2],
R. Gaitskell\footnotemark[7],
D. Jordan\footnotemark[1],
O. Kochetov\footnotemark[4],
S. Konovalov\footnotemark[3],
R. Kouzes\footnotemark[1],
H.S. Miley\footnotemark[1],
W.K. Pitts\footnotemark[1],
J.H. Reeves\footnotemark[1],
R.G.H. Robertson\footnotemark[6],
V. Sandukovsky\footnotemark[4],
E. Smith\footnotemark[1],
V. Stekhanov\footnotemark[3],
R.C. Thompson\footnotemark[1],
W. Tornow\footnotemark[8],$^{,}$\footnotemark[9],
V. Umatov\footnotemark[3],
R.A. Warner\footnotemark[1],
J. Webb\footnotemark[10],
J. Wilkerson\footnotemark[6],
and A. Young\footnotemark[11]}

\centerline{(The Majorana Collaboration)}

\vskip 1cm

The Majorana Experiment is a next-generation $^{76}$Ge
double-beta decay search. It will employ 500~kg of Ge,
isotopically enriched to 86\% in $^{76}$Ge, in the form of
$\sim $200 detectors in a close-packed array for high
granularity. Each crystal will be electronically segmented,
with each region fitted with pulse-shape
analysis electronics. A half-life sensitivity is predicted of
$\rm 4.2\times 10^{27}$~y or
$ \rm \langle m_{\nu} \rangle \leq 0.02-0.07$~eV,
depending on the nuclear matrix elements used to interpret the
data.

\newpage

\section{Introduction}
Neutrino oscillation experiments have produced ``smoking guns'' for
non-zero neutrino mass in the solar neutrino deficit\cite{bahcall}, in the
excess of $ p(\overline{\nu}_{e},e^{+})n $ reactions in the LSND
experiment\cite{mills}, and from the zenith-angle dependence of the
electron-to-muon event ratio in the Super-Kamiokande  (SK)
data\cite{sobel}
(see also \cite{fukuda,casper,ahlen}). The results of reactor neutrino
experiments\cite{apollonio}
constrain the disappearance of $ \overline{\nu}_{e} $ well enough
to imply that the SK data are dominated by $\nu_{\mu} \rightarrow
\nu_{\tau} \, (\overline{\nu}_{\mu} \rightarrow
\overline{\nu}_{\tau}) $ oscillations, with only a minimal
oscillation to electron-type neutrinos, since they show $
\overline{\nu}_{e}$'s do not oscillate as readily as required by
the SK data for $ \nu_{\mu}\rightarrow \nu_{e} $ or $
\overline{\nu}_{\mu}\rightarrow \overline{\nu}_{e} $ oscillations
to have a significant role.

While the interpretation of the SK data in terms of neutrino
oscillations is widely accepted, there were questions concerning
the interpretation of the LSND data as evidence of $
\overline{\nu}_{\mu} \rightarrow \overline{\nu}_{e} $
oscillations. Some doubt existed that the
standard solar model was accurate enough to support the
conclusion that there was really a deficit of solar neutrinos.
When the results of all solar neutrino experiments are
considered, there is no scenario in which these data are
compatible with the standard solar model unless the flux of
$\;\nu_{e}\; $ from the sun oscillates partially into other
$ \nu $-flavors to which the experiments are not sensitive.

On 17 July 2001, however, the Sudbury Neutrino Observatory (SNO)
collaboration settled this issue. They published results from the
direct measurement of the rate of the reaction $ d(\nu_{e},
e^{-})pp $ from solar neutrinos\cite{ahmad}. The neutrino flux implied by
these data was compared with that from neutral-current
neutrino-electron elastic scattering data from SK. It was
concluded that there is an active, non-electron-flavor neutrino
component in the solar neutrino flux, and that the total flux of
active neutrinos from the $ ^{8}$B branch is in close agreement
with the standard solar model of Bahcall and his co-workers\cite{bahcall}.
The standard solar model is thereby confirmed, and the case for
neutrino oscillations is now compelling.

A remaining question is that of the LSND indication of $
\overline{\nu}_{\mu} \rightarrow \overline{\nu}_{e} $
oscillations. All attempts to incorporate these results in the
same analysis with the solar neutrino and atmospheric neutrino
experiments fail in any scenario involving only three neutrino
flavors.

Accepting that now both solar and atmospheric neutrino experiments
give clear evidence for neutrino oscillations, there are only two
conclusions to be drawn from the LSND data. Either the excess
events from the reaction $ p(\overline{\nu}_{e}, e^{+})n $ in the
LSND is due to phenomena other than $ \overline{\nu}_{\mu}
\rightarrow \overline{\nu}_{e} $ oscillation, or that there must
exist a fourth generation of neutrinos ``sterile'' with respect to
``normal'' weak interactions\cite{ellis}. To insist on accepting one or
the other of these options at the present time is to accept an
unsubstantiated theoretical prejudice. This issue is still very
much an open one. The MiniBooNE experiment will settle this
issue; however, it will be a search for $ \nu_{\mu} \rightarrow
\nu_{e} $ ,not $ \overline{\nu}_{\mu} \rightarrow
\overline{\nu}_{e} $ as in LSND.

While an unambiguous interpretation of all of the neutrino oscillation
experiments is not yet possible, it is abundantly clear that neutrinos exhibit
properties not included in the standard model, namely mass and flavor
mixing. Accordingly, sensitive searches for neutrinoless double-beta decay
(0$\nu$~$\beta\beta$-decay) are more important than ever. Experiments
with kilogram quantities of germanium, isotopically enriched in
$ ^{76}$Ge, have thus far proven to be the most sensitive, specifically the
Heidelberg-Moscow\cite{baudis} and IGEX\cite{aalseth} experiments.
The lower limits in half-life sensitivities are:
$ 1.9 \times 10^{25}$~y\cite{baudis} and
$ 1.6 \times 10^{25}$~y\cite{aalseth}. A new generation of experiments
will be required to make significant improvements in sensitivity, as
discussed later.

Petcov and Smirnov\cite{petkov} showed that both MSW and vacuum
oscillation solutions of the solar neutrino problem can be
compatible with 0$\nu$~$\beta\beta$-decay driven by an
effective Majorana electron-neutrino mass in the range 0.1 to
1.0~eV.  The interpretation of all the neutrino oscillation data
together, as discussed later, implies a range that could be
between 5 and 10 times lower.  The exploration of such a range
will require next-generation experiments.  Some that are being
proposed are: CAMEO\cite{bellini}, CUORE\cite{allesandrello},
EXO\cite{danilov}, GENIUS\cite{klapdor},
Majorana\cite{braeckeleer}, and MOON\cite{ejiri}.

\section{Neutrinoless Double-Beta Decay}
Neutrinoless double-beta decay is a process by which two
neutrons in a nucleus beta decay by exchanging a virtual
Majorana neutrino, each emitting an electron. This violates lepton
number conservation $(\Delta l = 2)$. There are many reviews on
the subject\cite{primakoff,hax84,morales,fiorini}.

The decay rate for the process involving the exchange of a
Majorana neutrino can be written as follows:
\begin{equation}
\lambda^{0\nu}_{\beta\beta} = G^{0\nu}(E_0,Z)\langle m_\nu
\rangle^{2}|M^{0\nu}_f-( g_A /g_V )M^{0\nu}_{GT}|^{2}.
\label{eq:rate}
\end{equation}

In equation~\ref{eq:rate}, $ G^{0\nu}$ is the two-body phase-space factor
including coupling constants, $ M^{0\nu}_f $ and $ M^{0\nu}_{GT} $
are the Fermi and Gamow-Teller nuclear matrix elements
respectively, and $g_A$ and $g_V$ are the axial-vector and vector
relative weak coupling constants, respectively. The quantity
$\rm \langle m_{\nu} \rangle$
 is the effective Majorana neutrino mass given by:
\begin{equation}
\langle m_\nu \rangle \equiv \vert \sum^{2n}_{k=1} \lambda^{cp}_k (
U^L_{lk})^{2} m_k \vert ,
\label{eq:effmass}
\end{equation}
where $\lambda^{CP}_k$ is the $CP$ eigenvalue associated with the
$ k^{th} $ neutrino mass eigenstate  ($ \pm 1$ for $ CP$
conservation), $ U^L_{lk} $ is the $ (l,k )$ matrix element of the
transformation between flavor eigenstates $ | \nu_l\rangle $ and mass
eigenstates $ |\nu_k\rangle $ for left handed neutrinos,
\begin{equation}
|\nu_l \rangle = \sum U^L_{lk} | \nu_k \rangle,
\end{equation}
and $ m_k $ is the mass of the $ k^{th}$ neutrino mass eigenstate.

The effective Majorana neutrino mass, $\rm \langle m_{\nu} \rangle$, is
directly
derivable from the measured half-life of the decay as follows:
\begin{equation}
\langle m_{\nu} \rangle = m_e ( F_N T^{ 0\nu}_{ 1/2 } )^{ -1/2 } eV,
\end{equation} where
$ F_N \equiv G^{ 0\nu } | M^{ 0\nu}_f - ( g_A / g_V ) M^{
0\nu }_{ G T } |^{2}$, and $ m_e $ is the electron mass. This
quantity derives from nuclear structure calculations and is model
dependent.

The most sensitive experiments thus far utilize germanium detectors
isotopically enriched in $ ^{76}$Ge from 7.78\% abundance to
$ \sim$86\%. Germanium detector experiments were started with natural
abundance detectors by Fiorini, et al., in Milan\cite{fiorini}, evolving over
the years to the first experiments with small isotopically enriched detectors,
and finally to the two present multi-kilogram isotopically-enriched
$^{76}$Ge experiments: Heidelberg-Moscow\cite{baudis}
and IGEX\cite{aalseth}. Reference \cite{baudis} has about four times the
exposure as reference \cite{aalseth} with limits of similar magnitude. This
strongly implies that experiments with on the order of 100 moles of
$^{76}$Ge have reached the point of diminishing returns.

Where should the field of double-beta decay go from here? Suppose the
observed neutrino oscillations of atmospheric neutrinos and solar neutrinos
are considered. What probable range of $\rm \langle m_{\nu} \rangle$ is
implied? Would it be large enough to allow the direct observation of
0$\nu$~$\beta\beta$-decay? If so, what technique would be the best for a
possible discovery experiment?

\section{Theoretical Motivation and Probable Neutrino Scenarios}
The SK data imply maximal mixing of $ \nu_\mu $
with $ \nu_\tau $ with $ \delta m^2 _{23} \simeq 3\times10^{-3}$~eV$^2$.
The solar neutrino data from SK and from SNO also imply that
the small mixing angle solution to the solar neutrino problem
is disfavored, so that $ \delta m^2 $ (solar) $ \simeq (
10^{-5}-10^{-4})$~eV$^2$. Based on these interpretations, one
probable scenario for the neutrino mixing matrix has
the following approximate form:
\begin{equation}
{\Bigg( \begin{array}{c}
  {\nu_e} \\
  {\nu_{\mu}} \\
  {\nu_{\tau}}
\end{array}\Bigg) } \; = \;
{\Bigg(
\begin{array}{ccc}
  1/\sqrt{2} & 1/\sqrt{2} & 0 \\
  -1/\sqrt{2} & 1/\sqrt{2} & 1/\sqrt{2} \\
  1/\sqrt{2} & -1/\sqrt{2} & 1/\sqrt{2}
\end{array}
\Bigg)} {\Bigg( \begin{array}{c}
  {\nu_1} \\
  {\nu_{2}} \\
  {\nu_{3}}
\end{array}\Bigg) }
\end{equation}

The neutrino masses can be arranged in two hierarchical patterns in which
$ \delta m^2_{31} \simeq \delta m^2_{32}$
$\sim 3 \times 10^{-3}$~eV$^2 $, and
$ \delta m^2_{21} \sim ( 10^{-5} - 10^{-4})$~eV$^2$. With available data,
it is not possible to determine which hierarchy, $ m_3>m_1 (m_2)$, or
$ m_1(m_2) > m_3$, is the correct one, nor is the absolute value of any of
the mass eigenstates known.

The consideration of reactor neutrino and atmospheric neutrino data
together strongly implies that the atmospheric neutrino oscillations are very
dominantly $ \nu_\mu \rightarrow $
$\nu_\tau (\overline{\nu}_\mu \rightarrow \overline{\nu}_\tau )$, which
implies, as seen from equation~5, that $ \nu_e $ is very dominantly a
mixture of $ \nu_1 $ and $ \nu_2 $. In this case there will be one relative
$ CP $ phase, $ \epsilon $, and equation~\ref{eq:effmass} reduces to the
approximate form:
\begin{equation}
\rm \langle m_{\nu} \rangle= \frac{1}{2} ( m_1 + \epsilon m_2 ),
\end{equation}
where the large mixing angle solution of the solar
neutrino problem implies
\begin{equation}
( m^2_2 - m^2_1 ) = ( 10^{-5} - 10^{-4})\; \rm eV^2.
\end{equation}

Assuming that $ \epsilon \simeq + 1 $, and that neutrinos are Majorana
particles, then it is very probable that
0$\nu$~$\beta\beta$-decay is driven by an effective electron neutrino mass
between 0.01~eV and the present bound
from $^{76}$Ge experiments. Consider the relation
\begin{equation}
T^{0\nu}_{1/2} = {(\ln{2})Nt\over{c}},
\end{equation}
where $ N $ is the number of parent nuclei, $ t $ is the counting
time, and $ c $ is the maximum number of counts that can be
attributed to 0$\nu$~$\beta\beta$-decay. To improve the sensitivity to
$\rm \langle m_{\nu} \rangle$ by a factor of $ 10^{2}$ from the present
$\sim$1~eV to 0.01~eV, one must increase the quantity $ Nt / c $
by a factor of $ \rm 10^4$. The quantity $ N $ can feasibly be
increased by a factor of $\rm \sim 10^2$ over present experiments,
implying $ t/c $ must also be improved by a similar amount. Since the
present counting times are probably about a factor of 5 less than
a practical counting time, the background should be reduced by at least a
factor of 20 below present
levels. A further reduction is probably feasible and should be
pursued.

\section{The Majorana
$^{76}$Ge 0$\nu$~$\beta\beta$-Decay Experiment}
The Majorana experiment is proposed for a US deep underground
laboratory, and requires very little R\&D. It stands on the technical
shoulders of the IGEX experiment and other previous successful
double-beta decay and low-background experiments. Furthermore, new
segmented Ge detector technology has recently become commercially
available, while Pacific Northwest National Laboratory (PNNL)/University
of South Carolina (USC) researchers have developed new
pulse-shape discrimination techniques.

The IGEX experiment terminated with 117 mole-years of data with an
average background of 0.06~c/keV/kg/y in the later data
sets. The resulting half life is $\rm 1.6 \times 10^{25}$~y and its
implied bounds on $\rm \langle m_{\nu} \rangle$ are shown in Table~1.

\begin{figure}
\includegraphics[width=4in]{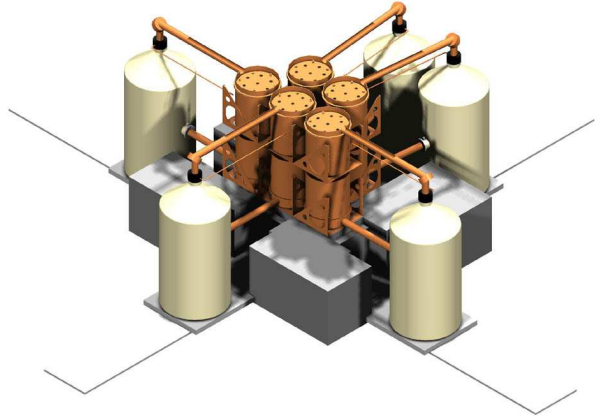}
\caption{ An artist's conception of the Majorana array of Ge
detectors using present cryostat technology. Other, more granular
arrays will also be investigated.}
\label{fig:phase3}
\end{figure}

The Majorana experiment represents a great increase in Ge mass over
IGEX with new segmented Ge detectors and the newest electronic systems
for pulse-shape discrimination. It is conceived to be 500~kg of Ge
detectors, fabricated from Ge isotopically enriched to 86\%  in $^{76}$Ge.
The typical detector size will be approximately 2.4~kg, requiring about 200
detectors. Below is presented the projected sensitivity for such an array,
shown in an artist's conception in Figure~\ref{fig:phase3}.

The array will have a fiducial mass of 500~kg, containing $ N
= 3.43 \times 10^{27} $ atoms of $ ^{76}$Ge; a counting time of
10 years, and an energy resolution of 3.0~keV FWHM. The
starting background assumed is that achieved by IGEX, prior to
pulse-shape discrimination, which was $ b_{0} = 0.2$~counts/keV/kg/y
at the 0$\nu$~$\beta\beta$-decay energy of 2039~keV. This
background was
achieved at 4000~mwe and was completely accounted for as the decay of
cosmogenic isotopes in the germanium\cite{bro95}. The decay of
$^{68}$Ge (270.8-day half-life) and $^{60}$Co (5.7-year half-life) over the
10-year counting period, combined with estimates of reduced
above-ground cosmogenic activation and additional decay during detector
construction, reduces this background by an overall factor of
about twenty, compared to the IGEX background mentioned. This reduces
the expected
background rate to 0.01~c/keV/kg/y.

The optimum energy window for a simple Poisson analysis is $2.8\sigma $
of the Gaussian peak (83.8\%), or 3.57~keV. The total number of
background counts in this window around 2039~keV over a 10-year
period would then be
\begin{equation}
b \cdot \Delta E \cdot M \cdot t = (0.01 \;\rm c/keV/kg/y)
(3.57 \; keV) (5000 \; kg \times y) = 178
\end{equation}
prior to data cuts.

\begin{figure}
\includegraphics[angle=-90,width=4in]{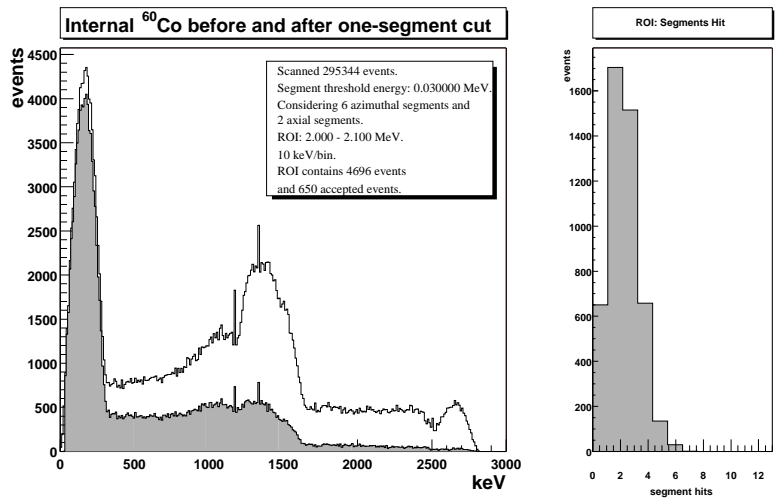}
\caption{Monte Carlo simulation of internal $^{60}$Co background. The
plot on the left shows the spectrum before and after a one-segment-only
cut is applied. The plot on the right displays a histogram of the number of
segments in which energy was deposited for internal $^{60}$Co
background events falling in a 2.0--2.1~ MeV region-of-interest.}
\label{fig:segmc}
\end{figure}

Above-ground experiments by the PNNL/USC collaborators demonstrate
that new pulse-shape discrimination methods have an acceptance fraction
for internal cosmogenic backgrounds of no more than 0.265. Monte Carlo
simulations demonstrate detector segmentation will allow an
additional cut with an acceptance fraction for internal cosmogenic
backgrounds of no more than 0.138. An example of the effect
of segmentation on internal cosmogenic background in the 2~MeV
region-of-interest is shown in Figure~\ref{fig:segmc}.
Applying both of these orthogonal cuts is expected to
reduce the background to 6.5 remaining events. If the background
in the region is featureless in energy, a simple Poisson analysis
yields a limit of 6 events as the number of 0$\nu$~$\beta\beta$-decay
events consistent with the background (90\%~CL).

The two cuts discussed do not have 100\% efficiency for accepting 0$\nu$
~$\beta\beta$-decay events. The pulse-shape discrimination has a
measured single-site-event acceptance fraction of 0.802. The
detector-segmentation cut has a calculated single-site-event acceptance
fraction of 0.907. The total counting efficiency is then 0.727. Accordingly
the sensitivity is projected as:
\begin{equation}
T^{o\nu}_{1/2}(sens)\; \simeq \; \frac{ (0.727)(.693)(3.43 \times
10^{27})\;10 }{6} \simeq 4.2 \times 10^{27} \rm \; y.
\end{equation}
This analysis does not account for the fact that the close-packed
granularity of $ \sim $200 crystals will allow further cuts against
multiply-scattered background events. The background computed above is
therefore conservative. The resulting upper bounds on
$\rm \langle m_{\nu} \rangle$ are shown in Table~\ref{tab:ff} and,
discarding an unfavored QRPA form factor, range from 0.02~eV to
0.07~eV.
\begin{table}[htb!]
\caption{Nuclear structure factor $F_N$ and Majorana neutrino mass
parameters $\rm \langle m_{\nu} \rangle$  for
$T^{0\nu}_{1/2}= 1.6\times 10^{25}$~y
and for $ 4.2 \times 10^{27}$~y.}
\begin{center}
\begin{tabular}{crcc}
  \hline
  $F_N$,yr$^{-1}$& Model [reference]
&  $\rm \langle m_{\nu} \rangle$ (eV) &
$\rm \langle m_{\nu} \rangle$ (eV)\\ \hline
    $1.56\times 10^{-13}$ & Shell Model\cite{haxton} & 0.32 & 0.020
\\
  $9.67\times 10^{-15}$ & QRPA\cite{vogel} & 1.3 & 0.080 \\
  $1.21\times 10^{-13}$ & QRPA\cite{civitarese}  & 0.37  & 0.023\\
  $1.12\times 10^{-13}$ & QRPA\cite{muto} &  0.38 & 0.024 \\
  $1.41\times 10^{-11}$ & Shell Model\cite{caurier} & 1.08 & 0.067
\label{tab:ff}
\end{tabular}
\end{center}
\end{table}
\vskip0.3cm



\section{Recent Progress in Ge Detector Technology}
Majorana is not simply a volume expansion of IGEX. It must have superior
background rejection. Because it has been conclusively shown that the
limiting background in at least some previous experiments has been
cosmogenic activation of the germanium itself, it is necessary to mitigate
those background sources. Cosmogenic activity fortunately has certain
factors which discriminate it from the signal of interest. For example, while
0$\nu$~$\beta\beta$-decay would deposit 2~MeV between two electrons
in a small, perhaps 1~mm$^3$ volume, internal $^{60}$Co decay deposits
about 318~keV (endpoint) in beta energy near the decaying atom, while
simultaneous 1173~keV and 1332~keV gammas can deposit energy
elsewhere in the crystal, most probably both in more than one location, for
a total energy capable of reaching the 2039~keV region-of-interest. A
similar situation exists for internal $^{68}$Ge decay. Thus
deposition-location multiplicity distinguishes double-beta decay from the
important long lived cosmogenics in germanium. Isotopes such as
$^{56}$Co, $^{57}$Co, $^{58}$Co, and $^{68}$Ge are produced at a
rate of roughly 1 atom per day per kilogram on the earth's
surface\cite{bro95}. Only $^{60}$Co and $^{68}$Ge have both the
energy and half-life to be of concern.

To pursue the multiplicity parameter, two approaches are possible. First,
the detector current pulse shape carries with it the record of energy
deposition along the electric field lines in the crystal; crudely speaking, the
radial dimension of cylindrical detectors. This information may be
exploited through pulse-shape discrimination. Second, the electrical
contacts of the detector may be divided to produce independent regions of
charge collection.

\begin{figure}[htb]
\begin{center}
\includegraphics[width=3in]{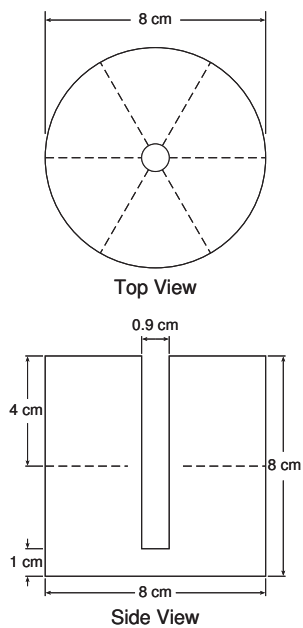}
\caption{The geometry of the segmented detectors planned for the
Majorana experiment.}
\label{fig:segdet}
\end{center}
\end{figure}

By segmenting the inner contact into two (axial) parts
and the outer contact into 6 (azimuthal) parts, as shown in
Figure~\ref{fig:segdet}, multiplicity data can be obtained.  The
Monte-Carlo simulation data set shown in Figure~\ref{fig:segmc} is based
on this configuration and shows that internal
highly-multiple backgrounds like $^{60}$Co can be strongly suppressed at
2038 keV. The internal $^{60}$Co modeled in the figure is produced by
cosmic-ray neutrons during the preparation of the detector, accumulating
after the last crystal-growth step. Its elimination by segmentation and pulse
shape discrimination is crucial. Beyond this simple segmentation cut, it may
be possible to use
the signals derived from segments seeing no net charge, adjacent to a
segment seeing net charge, to locate a single-site deposition in the axial and
azimuthal coordinates of the crystal or to distinguish a single-site
deposition from a multiple one. This possible technique is not currently
included in
sensitivity calculations. Intrinsic n-type Ge detectors with segmented
electrical contacts  are now available commercially. After Monte Carlo
studies and discussions with detector manufacturers, the general
configuration shown in Figure~\ref{fig:segdet} is believed to be optimal for
balancing background reduction, cost and production efficiency.

The development of the pulse-shape discrimination technique used in the
sensitivity calculation arose from earlier work comparing experimental
pulses with completely simulated ones. After repeated trials with fully
simulated pulses, this early method was set aside due to the unpredictable
effect of preamplifier noise levels and the unacceptable sensitivity of the
resulting cuts to the detector and charge-collection model. For example,
the parameters needed to correctly simulate pulses include the
space-charge density remaining after crystal depletion. This is most often
poorly known and represents essentially a set of free parameters for the
simulation. The operating voltage of the crystal, which may require
variation during the crystal lifetime, is also critically important to the fidelity
of simulated pulse formation. Thus, any system built upon the accuracy of
predicted or modeled pulse-shapes was found less than optimum for
cutting efficacy and fragile with respect to long-term maintenance.

A much-superior pulse-shape discrimination method was constructed using
a set of parameters calculated from each pulse. The parameters for a single
pulse are compared to a distribution collected from a calibration source of
interest. Those which conform to a large degree are kept; those which do
not are discarded. The acceptance ratio is adjusted during pulse
post-processing to optimize cut efficacy. The strength of this
approach is that no set of simulated pulses need be maintained; no space
charges need be known. It is only required that calibration data be obtained
before and after any voltage change, a procedure which is required in any
case to periodically check detector gain. To summarize, new
techniques depend entirely on experimental calibration and
do not utilize pulse-shape libraries. The ability of these techniques to be
easily calibrated for individual detectors makes them practical for large
detector arrays.

This development was in large part due to a shift in thinking about
digitization, earlier digitization had been done with an 8-bit digitizer
sampling as frequently as 500 MHz. Theory held that by oversampling, the
energy resolution afforded by analog acquisition could be achieved, which
is well-matched to 14-bits. However, measurement of actual preamplifier
bandwidth showed that there is no more than 25~MHz of information in the
signal. Thus, emerging hardware for 40-MHz 12-bit acquisition, such as the
Digital Gamma Finder (DGF) produced by X-Ray Instrumentation
Associates (XIA)\cite{nukl}, allows both a high-resolution energy value
and a high-fidelity waveform to be acquired in a single, lower cost, and
highly stable electronics chain. After acquiring a detector preamplifier
pulse, all subsequent operations on the signal are performed digitally. The
particular unit used in pulse-shape discrimination development is the
4-channel DFG-4C, developed and manufactured by XIA.

The DGF-4C has 4 independent, 40 MHz, 12-bit analog-to-digital
converters (ADCs). The ADCs are followed by First-in, First-out (FIFO)
buffers capable of storing 1024 ADC values for a single event. In parallel
with each FIFO is a programmable digital filter and trigger logic. The digital
filter and trigger logic for each channel is combined into a single Field
Programmable Gate Array (FPGA). Analog input data are continuously
digitized and processed at 40 MHz.

\begin{figure}[htb]
\begin{center}
\includegraphics[width=\textwidth]
{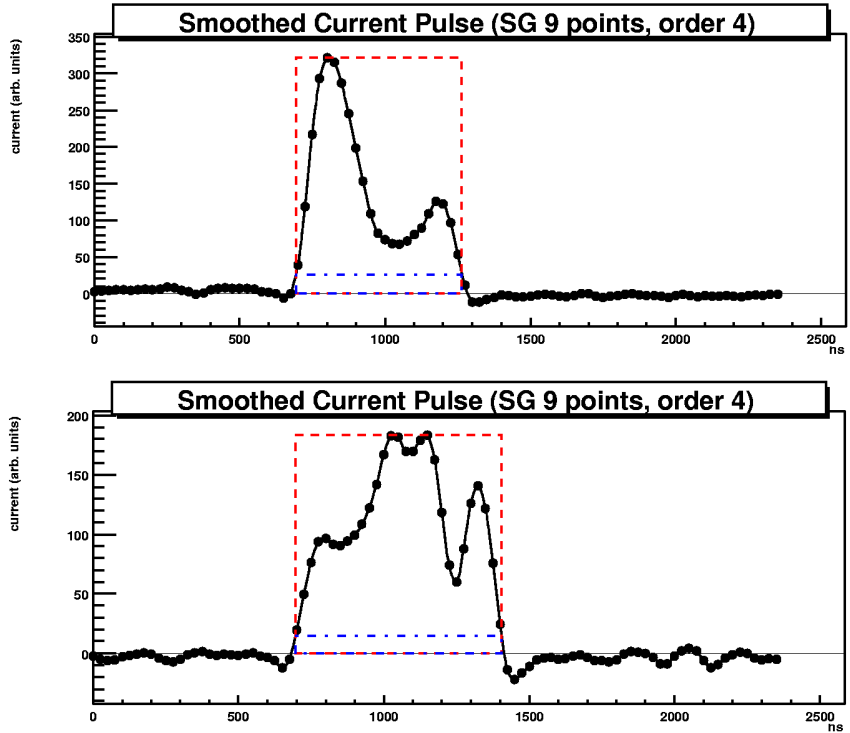}
\end{center}
\caption{Reconstructed current signal from a 1592-keV double-escape-%
peak event (top) and a 1620.6-keV full-energy gamma peak event (bottom).
Horizontal scale spans 2500~ns.}
\label{fig:psd20g}
\end{figure}

Experimental example pulses are shown
in Figure~\ref{fig:psd20g}.
An example single-site event from the
1592~keV double-escape peak of the $^{208}$Tl 2615~keV line is shown
as the top signal. The bottom signal is an example multi-site pulse from the
full-energy peak of the $^{212}$Bi line at 1620~keV.

The DGF-4C is a smart filter of incoming pulses. If, for example, a
signal has a pulse-width incompatible with the usual collection time of
200-1000~ns, or is oscillatory (like microphonic noise), the trigger filter can
be programmed to reject it. This feature can be used to allow the very low
energy thresholds required in Dark Matter searches as well as eliminating a
broad spectrum of very-low-rate artificial pulses from high voltage leaks
and electromagnetic interference that can appear in the energy region of
0$\nu$~$\beta\beta$-decay.

Calibration for single-site event pulses is trivially accomplished by
collecting pulses from thorium ore; the 2614.47-keV gamma ray from
$^{208}$Tl produces a largely single-site double-escape peak at
1592.47~keV. The PSD discriminator is then calibrated to the properties of
the double-escape peak A slightly improved double-escape peak can be
made from the $^{26}$Al gamma ray of 2938.22-keV. The double-escape
appears at 1916.22~keV, only about 120~keV away from the expected
region of interest for 0$\nu$~$\beta\beta$-decay.
The obvious and direct use of pulse-shape discrimination and segmentation
is the rejection of cosmogenic pulses in the germanium itself. However, the
approach should be also effective on gamma rays from the shielding and
structural materials.

The effects as background of neutrons of both high energy (cosmic muon
generated) and low energy (fission and ($\rm \alpha$,n) from rock) are
under consideration as well. The segmentation and granularity of the
detectors will provide some protection from this lower-order background.
These neutrons could also produce other unwanted activities. For instance,
$^{3}$H and $^{14}$C can be produced in nitrogen from high and low
energy neutrons, respectively. Fortunately, Majorana detectors will not be
surrounded by nitrogen at high density.

In conclusion, the Majorana project has been designed in a compact,
modular way such that it can be built and operated with high confidence in
the approach and the technology. The initial years of construction will
allow alternate cooling methods to be employed if they have an advantage
and should they be shown to overcome long-term concerns due to surface
contamination, muon-induced ions, and diffusion. The technology
supporting Majorana signal-processing-based background rejection is
ready and shows promise of future improvements. The technology of
radiopure copper production has improved steadily since the last IGEX
system, in which copper support materials played no measurable
radiological role. Finally, those who enrich germanium have also expressed
their enthusiasm and support for the project. Thus, the Majorana project is
poised ready to begin operations toward determining the effective
Majorana mass of the electron neutrino.


\begin{thebibliography}{99}
\bibitem{bahcall}J.N. Bahcall, Centennial Lecture, Neutrino-2000,
Nucl. Phys. B (Proc. Suppl.) {\bf 91}, 9 (2001).
\bibitem{mills}G.B. Mills, Nucl. Phys. B (Proc. Suppl.) {\bf 91}, 198
(2001).
\bibitem{sobel}H. Sobel (for the Super-Kamiokande Collaboration), Nucl.
Phys. B (Proc. Suppl.) {\bf 91}, 127 (2001); Y. Fakuda, {\it et
al.}, Phys. Rev. Lett. {\bf 81}, 1562 (1998); {\bf 82}, 1810
(1999).
\bibitem{fukuda}Y. Fukuda, {\it et al.}, Phys. Lett. {\bf B 335},
237 (1994).
\bibitem{casper}D. Casper, {\it et al.}, Phys. Rev. Lett. {\bf 66},
2561 (1991); R. Becker-Szendy, {\it et al.}, Phys. Rev. {\bf D 46},
3720 (1992).
\bibitem{ahlen}S. Ahlen, {\it et al.}, Phys. Rev. Lett. {\bf 72}, 608 (1994).
\bibitem{apollonio}M. Apollonio, {\it et al.}, Phys. Lett. {\bf B 420}, 397
(1998); Phys. Lett. {\bf B 466}, 415 (1999). The CHOOZ
Collaboration, and F. Boehm, {\it et al.}, Nucl. Phys. {\bf 91},
91 (2001). The Palo Verde Collaboration.
\bibitem{ahmad}Q.R. Ahmad, {\it et al.}, Phys. Rev. Lett. {\bf
87}, 071301-1 (2001).
\bibitem{ellis}J. Ellis (Summary Talk) Neutrino-2000, Nucl. Phys. B (Proc.
Suppl.) {\bf 91}, 503 (2001).
\bibitem{baudis}L. Baudis, {\it et al.}, Phys. Rev. Lett. {\bf 83},
41 (1999); see also F.T. Avignone III, C.E. Aalseth, and
R.L. Brodzinski, Phys. Rev. Lett. {\bf 85}, 465 (2000). (The value
of $ 1.9 \; \times \; 10^{25} $ years is the new official
Heidelberg Value Presented by H.V. Klapdor-Kleingrothaus at
NANP-2001, June 2001, Dubna, Russia).
\bibitem{aalseth}C.E. Aalseth, {\it et al.}, (IGEX Collaboration),
Yad. Phys. {\bf 63}, 1299 (2000); D. Gonzales, Nucl. Phys. B
(Proc, Suppl.) {\bf 87}, 278 (2000).
\bibitem{petkov}S.T. Petcov and A.Yu. Smirnov, Phys. Lett. {\bf B 322},
109 (1994).
\bibitem{bellini}G. Bellini, {\it et al.}, IFN Preprint arXiv:
nucl.-ex/0007012, 11 July 2000, SIS-Pubblicazioni dei Laboratori
di Frascati.
\bibitem{allesandrello}A. Allesandrello, {\it et al.}, Phys. Lett. {\bf B 420},
109 (1998); Nucl. Phys. {\bf B 87}, 78 (2000).
\bibitem{danilov}M. Danilov, {\it et al.}, Phys. Lett. {\bf B 480},
12 (2000).
\bibitem{klapdor}H.V. Klapdor-Kleingrothaus, J. Helmig, and M. Hirsch,
J.Phys. G: Nucl. Part. Phys. {\bf 24}, 483 (1998).
\bibitem{braeckeleer}L. De Braeckeleer (for the Majorana
Collaboration), Proc. Carolin Conf. on Neutrino Physics (Columbia
SC USA, 10-12 March 2000), eds. J.N. Bahcall, W.C. Haxton,
K. Kubodera, and C.P. Poole, World Scientific.
\bibitem{ejiri}H. Ejiri, {\it et al.}, Phys. Rev. Lett. {\bf 85},
2917 (2000).
\bibitem{primakoff}H. Primakoff and S.P. Rosen, Rep. Prog. Phys.
{\bf 22}, 121 (1959); Phys. Rev. {\bf 184}, 1925 (1969).
\bibitem{hax84}W.C. Haxton and G.J. Stevenson Jr., Prog. Part. Nucl.
Phys. {\bf 12}, 409 (1984); F.T. Avignone III and R.L. Brodzinski,
Prog. Part. Nucl. Phys. {\bf 21}, 99 (1988); M. Moe and P. Vogel,
Ann. Rev. Nucl. Part. Sci {\bf 44}, 247 (1994).
\bibitem{morales}A. Morales, Nucl. Phys. B (Proc. Suppl.) {\bf
77}, 335 (1999); H. Ejiri, Phys. Rept. C (2000), Int. J. Mod.
Phys. E6, 1 (1997); V. Tretyak and Y. Zdesenko, At. Data, Nucl.
Data Tables {\bf 61}, 43 (1995).
\bibitem{fiorini}E. Fiorini {\it et al.}, Phys. Lett. {\bf 25 B}, 602
(1967); Lett. Nuovo Cimento {\bf 3}, 149 (1970).
\bibitem{bah95}J.N. Bahcall and M.H. Pinsonneault, Rev. Mod. Phys.
{\bf 67}, 781 (1995). For an update on uncertainties in the
models see J.N. Bahcall, S. Basu, and M.H. Pinsonneault, Phys. Lett.
{\bf B 433}, 1 (1998).
\bibitem{bro95}
R.~L. Brodzinski, H.~S. Miley, J.~H. Reeves, and F.~T. Avignone, III.
\newblock Low-background germanium spectrometery -- the bottom line
three years
  later.
\newblock {\em Journal of Radioanalytical and Nuclear Chemistry},
  193(1):61--70, 1995.
\bibitem{haxton}W.C. Haxton and G.T. Stephenson Jr., Prog. Part.
Nucl. Phys. {\bf 12}, 409 (1984).
\bibitem{vogel}P. Vogel and M.R. Zirnbauer, Phys. Rev. Let. {\bf
57}, 3148 (1986); J. Engel, P. Vogel, and M.R. Zirnbauer, Phys. Rev.
{\bf C 37}, 731 (1988).
\bibitem{civitarese}O. Civitarese, A. Faessler, and T. Tomoda, Phys.
Lett. {\bf B 194}, 11 (1987); T. Tomoda and A. Faessler, Phys. Lett.
{\bf B 199}, 475 (1987).
\bibitem{muto}K. Muto, E. Bender, and H.V. Klapdor, Z.Phys. {\bf A
177}, 334 (1989). The matrix elements used in Table 1 are from
A. Standt, K. Muto, and H.V. Klapdor, Europhys. Lett. {\bf 13}, 31
(1990).
\bibitem{caurier}E. Caurier, {\it et al.}, Phys. Rev. Lett. {\bf
77}, 1954 (1996).
\bibitem{nukl}B. Hubbard-Nelson, M. Momayezi and W.K. Warburton,
Nucl. Instr. and Meth. in Physics Research {\bf A
422}, 411 (1999).

\end{thebibliography}
\end{document}